\documentclass{amsart}
\usepackage[foot]{amsaddr}
\pdfoutput=1

\usepackage{amssymb}
\usepackage{mathtools}
\usepackage{amscd}
\usepackage[mathscr]{eucal}
\usepackage{enumitem}
\usepackage[hmargin=1in,vmargin=1in]{geometry}

\usepackage[pdftex]{graphicx, color}

\usepackage[roman]{sublabel}

\usepackage[labelfont=bf]{caption}
\usepackage{subcaption}

\usepackage{cite}
\usepackage{comment}

\definecolor{link_blue}{rgb}{0,0,0.97}
\definecolor{cite_red}{rgb}{0.7,0,0}
\definecolor{dark_green}{rgb}{0,.6,0}
\definecolor{purple}{rgb}{0.7,0.25,0.7}
\definecolor{yoda}{rgb}{0.7,0,0}

\usepackage[colorlinks=true, citecolor=cite_red, linkcolor=link_blue]{hyperref}

\usepackage{ottmacros_cell_v1}
\DeclareMathOperator{\Cov}{Cov}

\newcommand{\lrate}[3]{\lambda^{#1}_{#2 \to #3}}

\makeatletter
\newcommand*{\da@rightarrow}{\mathchar"0\hexnumber@\symAMSa 4B }
\newcommand*{\da@leftarrow}{\mathchar"0\hexnumber@\symAMSa 4C }
\newcommand*{\xdashrightarrow}[2][]{%
  \mathrel{%
    \mathpalette{\da@xarrow{#1}{#2}{}\da@rightarrow{\,}{}}{}%
  }%
}
\newcommand{\xdashleftarrow}[2][]{%
  \mathrel{%
    \mathpalette{\da@xarrow{#1}{#2}\da@leftarrow{}{}{\,}}{}%
  }%
}
\newcommand*{\da@xarrow}[7]{%
  \sbox0{$\ifx#7\scriptstyle\scriptscriptstyle\else\scriptstyle\fi#5#1#6\m@th$}%
  \sbox2{$\ifx#7\scriptstyle\scriptscriptstyle\else\scriptstyle\fi#5#2#6\m@th$}%
  \sbox4{$#7\dabar@\m@th$}%
  \dimen@=\wd0 %
  \ifdim\wd2 >\dimen@
    \dimen@=\wd2 %
  \fi
  \count@=2 %
  \def\da@bars{\dabar@\dabar@}%
  \@whiledim\count@\wd4<\dimen@\do{%
    \advance\count@\@ne
    \expandafter\def\expandafter\da@bars\expandafter{%
      \da@bars
      \dabar@ 
    }%
  }%
  \mathrel{#3}%
  \mathrel{%
    \mathop{\da@bars}\limits
    \ifx\\#1\\%
    \else
      _{\copy0}%
    \fi
    \ifx\\#2\\%
    \else
      ^{\copy2}%
    \fi
  }%
  \mathrel{#4}%
}
\makeatother

\theoremstyle{plain}

\theoremstyle{definition}

\newtheorem*{acknowledgments}{Acknowledgments}

\begin{document}

\title[Effects of cell cycle noise]{Effects of cell cycle noise on excitable gene circuits}

\author{Alan Veliz-Cuba$^{1}$}
\author{Chinmaya Gupta$^{2}$}
\author{Matthew R.\ Bennett$^{3}$}
\author{Kre\v{s}imir Josi\'{c}$^{2,3,4}$}
\author{William Ott$^{2}$}

\address{$^{1}$Department of Mathematics, University of Dayton}
\address{$^{2}$Department of Mathematics, University of Houston}
\address{$^{3}$Department of Biosciences and Department of Bioengineering, Rice University}
\address{$^{4}$Department of Biology and Biochemistry, University of Houston}



\date{\today}

\begin{abstract}
We assess the impact of cell cycle noise on gene circuit dynamics.
For bistable genetic switches and excitable circuits, we find that transitions between metastable states most likely occur just after cell division and that this {\itshape concentration effect} intensifies in the presence of transcriptional delay.
We explain this concentration effect with a $3$-states  stochastic model.
For genetic oscillators, we quantify the temporal correlations between daughter cells induced by cell division.
Temporal correlations must be captured properly in order to accurately quantify noise sources within gene networks.
\end{abstract}

\maketitle

\thispagestyle{empty}

\section{Introduction}
\label{s:intro}

\renewcommand{\thefootnote}{\fnsymbol{footnote}} 
\footnotetext{\emph{PACS codes.} 02.50.Ey, 87.10.Mn, 87.17.Ee, 87.18.Cf, 87.18.Tt}     
\renewcommand{\thefootnote}{\arabic{footnote}}

Cellular noise and transcriptional delay shape the dynamics of genetic regulatory circuits.
Stochasticity in cellular processes has a variety of sources, ranging from low molecule numbers~\cite{Arkin01081998,Ozbudak2002,Elowitz2002,Becskei2005,St-Pierre30122008,Hensel2012},
to variability in the environment, metabolic processes, and available energy~\cite{Thattai2004,Hilfinger19072011,roberts2011, Zopf2013,Swain01102002,Raser2004, Volfson2006, Shahrezaei2008, Dunlop2008,Hilfinger19072011}.
Such fluctuations can drive a variety of dynamical phenomena, including oscillations~\cite{Turcotte14102008}, stochastic state-switching~\cite{NatGen_Acar:2008}, and pulsing~\cite{PCompBIo_GarciaBernardo:2013}.  Microbial and eukaryotic cells make use of such dynamics in probabilistic differentiation strategies to stochastically switch between gene expression states~\cite{ozbudak:2004},  and for transient cellular differentiation~\cite{Eldar-Elowitz-Functional-2010, Suel_2006, Davidson-Surette-Individuality-2008}.
%

How cell cycle noise shapes dynamics is only partially understood.
The cycle of cell growth and division results in a distinct noise pattern:
Intrinsic chemical reaction noise decreases as cells grow before abruptly jumping following cell division.
The partitioning of proteins and cellular machinery at division also induces  a temporally localized, random perturbation in the two daughter cells.
These perturbations are correlated, as a finite amount of cellular material is divided between the two descendant cells.  Such correlations can propagate across multiple generations within a lineage~\cite{veliz_noise:2015}.

We find that cell cycle noise can strongly impact the dynamics of bistable and excitable systems.
In both cases, transitions out of metastable states are concentrated within a short time interval just after cell division.
Interestingly, this effect intensifies as transcriptional delay (the time required for a regulator protein to form and signal its target promoter) increases.
We show that this concentration effect results primarily from the random partitioning of cellular material upon cell division, and explain the underlying mechanisms by extending a $3$-states reduced model introduced in~\cite{Gupta-stabilization-PRL:2013}.
For genetic oscillators, we find that cell cycle noise plays an important role in shaping temporal correlations along descendant lineages.
In particular, for the synthetic genetic oscillator described in~\cite{Stricker2008}, we show that temporal correlations between daughter cells decay significantly faster when the cell cycle is modeled explicitly.

In models of genetic networks the effects of cell growth are frequently described by a simple dilution term which does not capture the distinct temporal characteristics of cell cycle noise.
We conclude that in order to accurately describe gene circuit dynamics, such models should include both cell cycle noise and transcriptional delay.

\begin{figure}[ht]
\begin{center}
\includegraphics[scale=0.9]{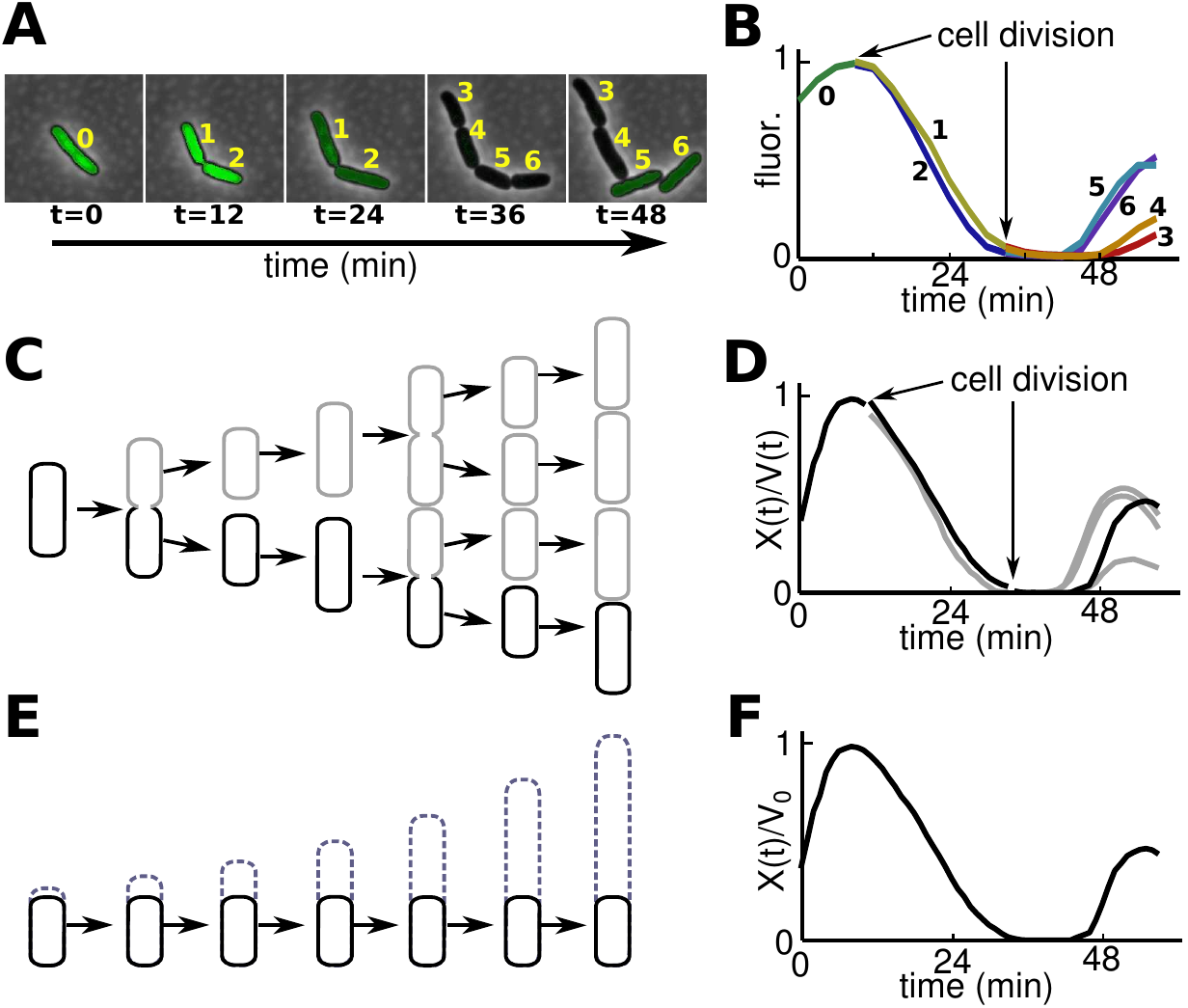}
\end{center}
\caption{
Different models of the cell cycle.
{\bfseries (A)}
A single cell gives rise to a lineage. 
Copies of a dynamic genetic network (in this case an oscillator, \cite{Stricker2008}) are passed to the daughter cells upon division.
Oscillations persist along the lineage.
{\bfseries (B)} 
Fluorescence traces recorded from different cells show how each trajectory branches into two at cell division~\cite{veliz_noise:2015}.
Random partitioning of cellular material at cell division initiates the branching process. 
{\bfseries (CD)}
When modeling cell growth and division explicitly, cell volume grows exponentially before being halved at the time of division.
The trajectory of a representative cell (bold in C) is shown in D.
{\bfseries (EF)}
In the dilution modeling framework, a single compartment that grows indefinitely represents the population.
A fixed subvolume of this compartment (bold in E) represents the average behavior of cells in the population.
}
\label{fig:cell_div_modeling}
\end{figure}

\section{How fluctuations depend on cell cycle phase for constitutive protein production}
\label{s:equilibrium}

We begin by examining the impact of cell cycle noise on simple constitutive protein production.
We find that at low to moderate protein numbers, cell division noise primarily determines 
how fluctuations in protein concentration depend on cell cycle phase.
Intrinsic noise plays a secondary role.

Constitutive production of a protein $P$ is described by the reaction
\begin{equation*}
\emptyset \xrightarrow{\psi (X)} P,
\end{equation*}
where $X$ denotes the number of proteins $P.$ The reaction rate, $\psi (X),$ is given by $\psi (X) = \Omega (t) f (X / \Omega (t))$, where $\Omega (t)$ denotes cell volume and $f$ is the reaction propensity function.
In terms of concentration, $x = X / \Omega (t)$, the reaction propensity is constant, $f(x) = \alpha $.

We simulate this system using a hybrid algorithm described by the augmented system
\begin{alignat*}{2}
&\emptyset \xrightarrow{\psi (X)} P & &\qquad \text{(simulate using a stochastic simulation algorithm)}
\\
&\frac{\mrm{d} \Omega }{\mrm{d} t} = \gamma \Omega & &\qquad \text{(cell division occurs when $\Omega = 2$)}.
\end{alignat*}
Here $\gamma = \ln (2)$ is the cell growth rate, and we measure time in units of cell cycle length.
We assume deterministic cell volume growth (from $\Omega = 1$ to $\Omega = 2$), and a fixed, volumetric threshold for cell division.
Upon division proteins are binomially partitioned between the daughter cells.
We use a (delay) stochastic simulation algorithm (dSSA) to simulate chemical reactions, with reaction rates depending on cell volume (See Fig.~\ref{fig:cell_div_modeling} for an illustration, and Section~\ref{s:modeling} for details).

\begin{figure}[ht]
\begin{center}
\includegraphics[scale=0.4]{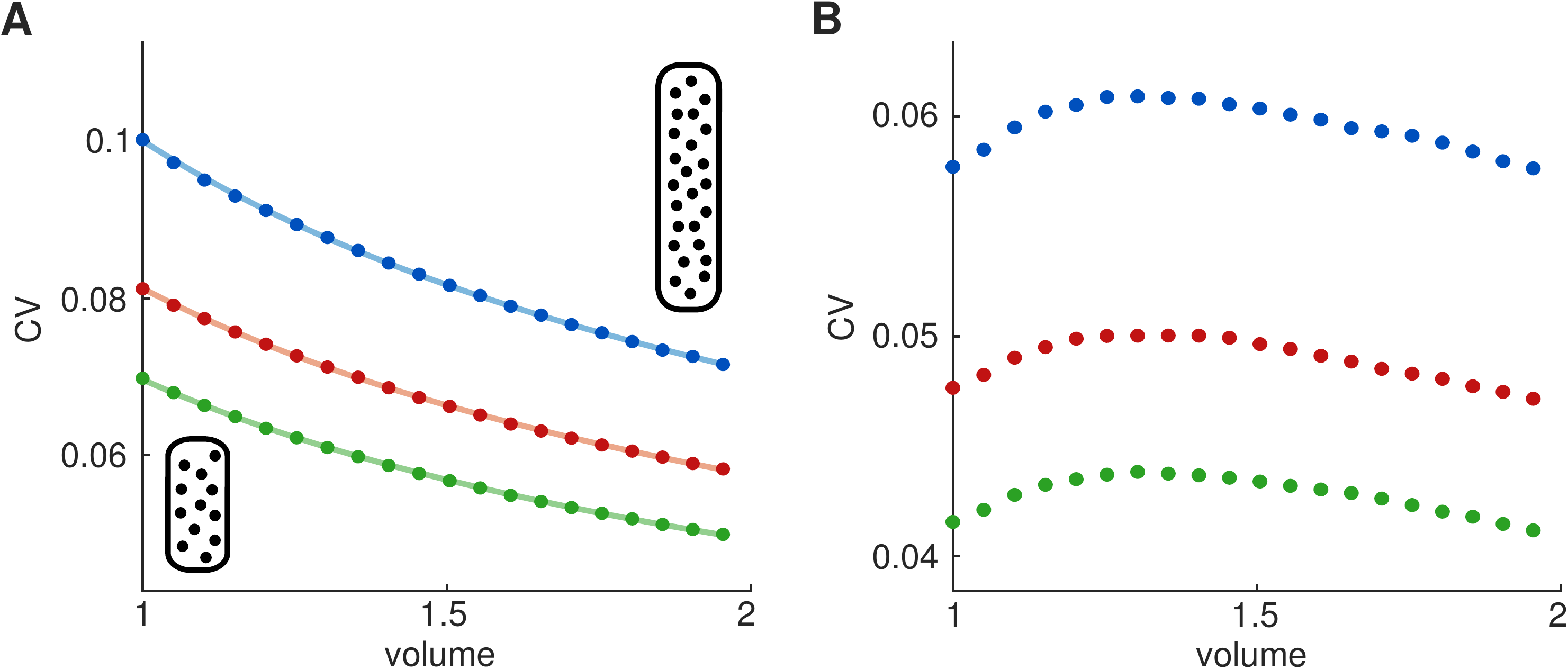}
\end{center}
\caption{
Effects of cell cycle noise on constitutive protein production for different values of $\langle x \rangle=\alpha/\gamma$ (blue: $\langle x \rangle=100$, red: $\langle x \rangle=150$, green: $\langle x \rangle=200$).
{\bfseries (A)} 
CV of protein concentration as a function of volume (filled circles).
The CV is highest early in the cell cycle and decreases as the cell grows. 
The solid curves show best fits of simulations to  functions proportional to $\Omega^{-1/2}$.
{\bfseries (B)}
Removing binomial partitioning noise reduces CV magnitude and significantly reduces the amount by which CV varies as a function of cell volume, the latter by roughly an order of magnitude.
}
\label{fig:time_dependent_noise}
\end{figure}

Fig.~\ref{fig:time_dependent_noise}A illustrates the combined effect of intrinsic chemical reaction noise and binomial partitioning noise on the coefficient of variation (CV) of $x$.
As a function of cell volume, the coefficient of variation $\sigma_{x} / \langle x \rangle $ scales as $\Omega^{-1/2}$.
In particular, the CV is highest at the beginning of the cell cycle and decreases monotonically until the next division.

Of the two noise sources, binomial partitioning noise is the major factor that determines the dependence of the 
CV on cell cycle phase.
Fig.~\ref{fig:time_dependent_noise}B illustrates the result of modifying the hybrid algorithm by dividing proteins evenly between daughter cells upon cell division, thus removing partitioning noise.
This results in a CV with a peak towards the middle of the cycle. As expected, removing  partitioning noise decreases CV magnitude. Importantly, we observe that the amount by which CV varies across the cell cycle decreases significantly, by roughly an order of magnitude.

Fig.~\ref{fig:time_dependent_noise} also shows that the dependence of the CV on cell cycle phase is independent of protein number. 
Changing the protein number essentially rescales the CV curve - shape is preserved.

\section{Bistable switches and excitable systems}
\label{s:metastability}

We computationally study the co-repressive toggle switch and a representative excitable system.
In both cases, we find that transitions out of metastable states concentrate in a small interval of time just after cell division.
Intuitively, this concentration effect results from the observation that stochastic fluctuations are maximal at the beginning of the cell cycle due to cell division (Fig.~\ref{fig:time_dependent_noise}A).
We show that the effect largely disappears when binomial partitioning noise is removed.
This happens intuitively because stochastic fluctuations are significantly more uniform across the cell cycle without partitioning noise (Fig.~\ref{fig:time_dependent_noise}B). 

The concentration effect intensifies with the addition of transcriptional delay due to a subtle interplay between delay and cell cycle noise.
We explain this intensification in Section~\ref{s:symbol} using a $3$-states stochastic model.

\subsection{An archetypal bistable system}

\begin{figure}[ht]
\begin{center}
\includegraphics[scale=0.84]{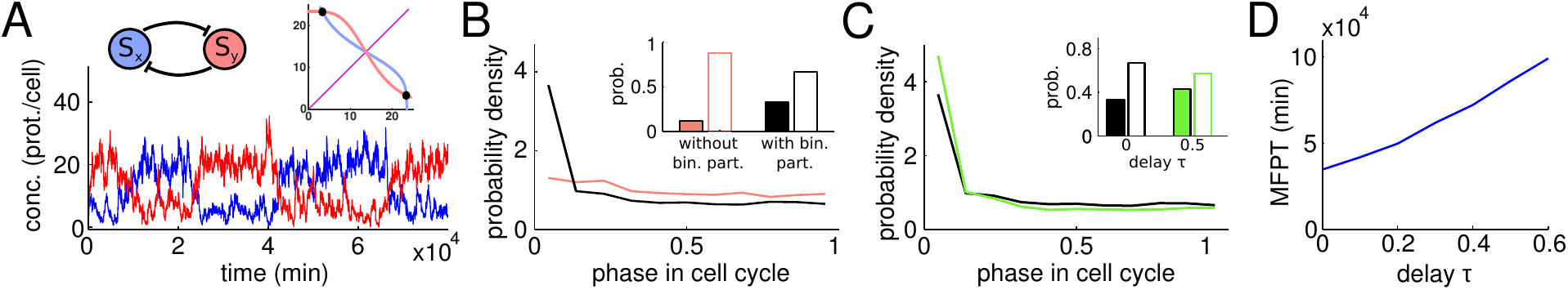}
\end{center}
\caption{Effects of cell growth and division on a co-repressive toggle switch. 
{\bfseries (A)} 
Network diagram and typical dynamics of the stochastic system displaying transitions between two metastable states.
The inset illustrates the phase portrait in the deterministic limit: nullclines (blue and red curves), stable steady states (solid circles), and unstable manifold of the saddle point (purple).
{\bfseries (B)} 
Probability density of transitions between metastable states within the cell cycle.
Transitions most likely occur right after cell division (black curve). 
Removing partitioning noise by equally dividing proteins upon cell division results in a much more uniform transition probability distribution (red curve).
Filled bars: $\mathrm{Prob} (t \leqs 0.1)$, empty bars: $\mathrm{Prob} (t>0.1)$.
{\bfseries (C)} 
Delay in protein production increases the probability that a transition occurs just after cell division.
Delay values: $\tau = 0$ (black), $\tau = 0.5$ (green).
Filled bars: $\mathrm{Prob} (t \leqs 0.1)$, empty bars: $\mathrm{Prob} (t>0.1)$.
{\bfseries (D)} 
Mean first passage time as a function of delay. 
As delay in protein production increases, the metastable states become more stable.
We multiply the birth propensity function $f$ by $e^{\gamma \tau }$ to ensure that the dynamics of the switch may be fairly compared as $\tau $ varies (see~\cite{obrien:2012} for an explanation).
Other parameter values: $\alpha = 16.25$, $k = 4$, $\beta = 15$, $\gamma = \ln (2)$.
}
\label{fig:toggle}
\end{figure}

The co-repressive toggle switch (Fig.~\ref{fig:toggle}A, inset) is an archetypal model of bistability described by the biochemical reaction network
\begin{subequations}
\label{e:coRT_network}
\begin{align}
&\emptyset \xdashrightarrow[\tau]{\psi (Y)} S_{X} 
\\
&\emptyset \xdashrightarrow[\tau]{\psi (X)} S_{Y}.
\end{align}
\end{subequations}
Here $X$ and $Y$ denote molecule numbers of the species $S_{X}$ and $S_{Y}$, respectively.
The dashed arrows indicate that the production reactions include a fixed delay, $\tau $.
Although we opt to simulate~\eqref{e:coRT_network} with fixed delay, we expect that our results hold for distributed delay as well.  
The reaction rate $\psi (\cdot )$ in this symmetric system is given by $\psi (\cdot ) = \Omega (t) f (\cdot / \Omega (t))$, where $f$ is the reaction propensity function
\begin{equation*}
f (\cdot ) = \frac{\alpha}{1 + (\cdot / \beta )^{k}}.
\end{equation*}

We simulate the co-repressive toggle switch using the hybrid algorithm described in Section~\ref{s:equilibrium}.
The production reactions for $S_{X}$ and $S_{Y}$ include delay, so we use the dSSA to simulate chemical reactions.
When cell division occurs, molecules of $S_{X}$ and $S_{Y}$ as well as the protein production queues for both molecular species are binomially partitioned between the two daughter cells.

In the $\Omega \to \infty $ limit, the co-repressive toggle is described by the delay reaction rate equations~\cite{Higham:2008:Modeling, Gupta_jcp2014, Brett-Galla-PRL-2013}
\begin{subequations}
\label{e:coRT_dRRE}
\begin{align}
\frac{\mrm{d} x}{\mrm{d} t} &= \frac{\alpha }{1 + (y(t - \tau ) / \beta )^{k}} - \gamma x
\\
\frac{\mrm{d} y}{\mrm{d} t} &= \frac{\alpha }{1 + (x(t - \tau ) / \beta )^{k}} - \gamma y,
\end{align}
\end{subequations}
where $x$ and $y$ denote the concentrations of $S_{X}$ and $S_{Y}$, respectively.
We choose parameters for which system~\eqref{e:coRT_dRRE} has two stable stationary points $(x_{l},y_{h})$ and $(x_{h}, y_{l})$ separated by an unstable manifold associated with a saddle equilibrium point $(x_{s}, y_{s})$ (Fig.~\ref{fig:toggle}A, inset).

When $\Omega < \infty $, the system is stochastic, and the stable stationary points $(x_{l}, y_{h})$ and $(x_{h}, y_{l})$ become metastable.
In this regime, a typical trajectory will spend most of its time near the metastable points, occasionally hopping from one to the other (Fig.~\ref{fig:toggle}A).

Fig.~\ref{fig:toggle} shows that cell cycle noise has a strong effect on the dynamics of the co-repressive toggle switch.
The black curves show the conditional probability density function (PDF) of transition times between metastable states within the cell cycle (conditioned on a transition having occurred) in the absence of delay ($\tau = 0$).
Such transitions most likely happen early in the cell cycle - a concentration effect.

To isolate the cause of this spike, we examine the impact of partitioning noise at cell division and intrinsic biochemical noise throughout the cell cycle.
If we divide proteins evenly between daughter cells at division, thereby removing partitioning noise, the spike in transition probability right after cell division disappears (Fig.~\ref{fig:toggle}B, red conditional PDF).
Instead, the probability of a transition decreases gradually with cell cycle phase.
We attribute this gradual decrease to the decrease in intrinsic noise that accompanies increasing cell volume. 

The concentration effect is robust with respect to the number of proteins in the system.
Transitions between metastable states become less frequent as protein numbers increase because noise levels decrease.
However, the PDFs exhibiting the concentration effect are conditioned on a transition having occurred.
Their shape therefore depends on the shape, and not the magnitude, of cell cycle noise.
Fig.~\ref{fig:time_dependent_noise} shows that cell cycle noise shape does not depend on protein number.

The concentration effect intensifies with the addition of transcriptional delay.
Fig.~\ref{fig:toggle}C  shows the effect when delay is positive ($\tau = 0.5$ (green), and 
$\tau = 0$ (black) for comparison).
Transitions most likely occur near the beginning of the cell cycle for $\tau = 0$ and this effect intensifies (probability mass further concentrates near the moment of cell division) for $\tau = 0.5$. 
We will show in Section~\ref{s:symbol} that this effect is caused by a subtle interplay between reaction and partitioning noise.

We have shown previously that  transcriptional delay stabilizes bistable genetic networks~\cite{Gupta-stabilization-PRL:2013}.
For a variety of circuits that exhibit metastability, increasing the delay dramatically increases mean residence times near metastable states. However, in this original study we only included a dilution term. Fig.~\ref{fig:toggle}D shows that this stabilization effect persists when cell growth and division are modeled explicitly.

\subsection{Excitable dynamics}

\begin{figure}[ht]
\begin{center}
\includegraphics[scale=0.84]{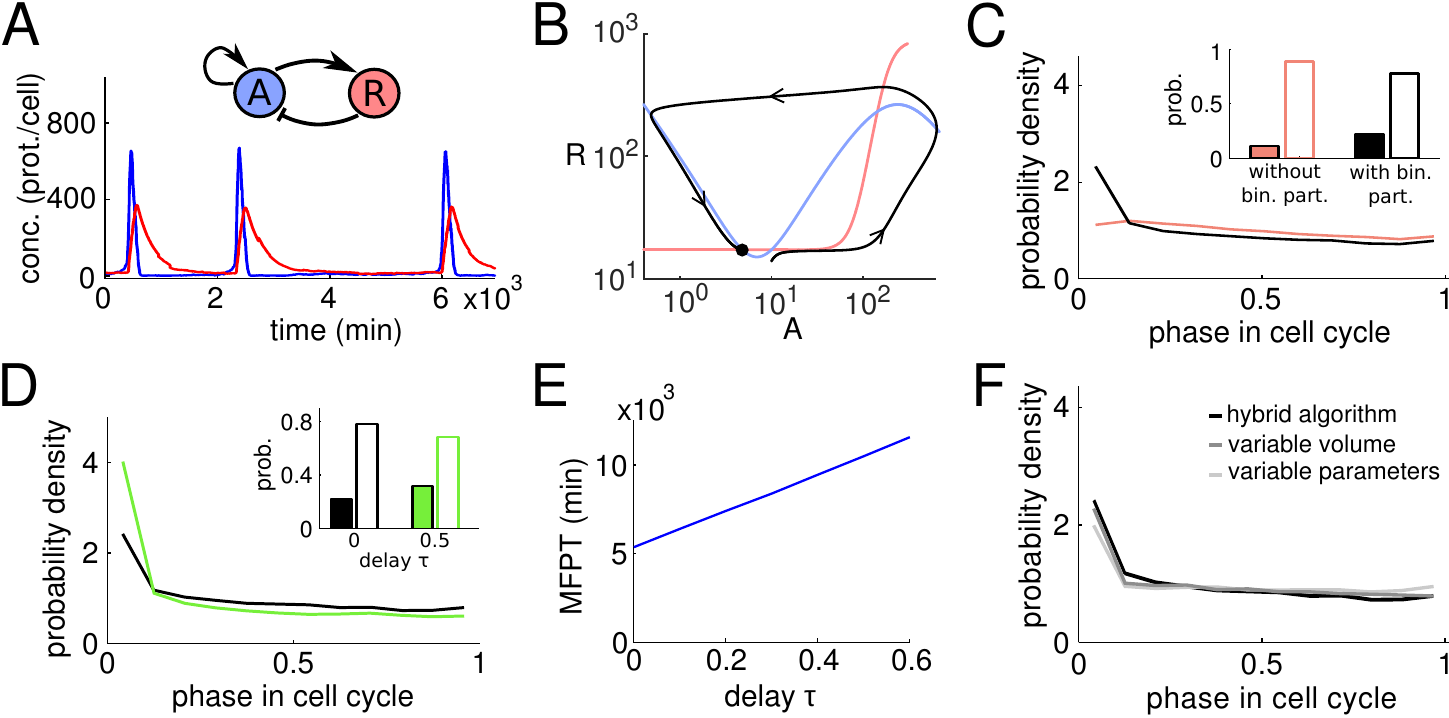}
\end{center}
\caption{Effects of cell growth and division on an excitable system.
{\bfseries (A)} 
Network diagram and typical dynamics exhibiting  sporadic pulses. 
{\bfseries (B)}
Phase portrait in the deterministic limit:
Nullclines (blue and red curves), the stable steady state (solid circle), and a deterministic trajectory that makes an excursion around the unstable points before returning to the stable steady state (black curve).
{\bfseries (C)} 
The probability density of pulse times within the cell cycle shows that pulses are more likely soon after cell division (black curve). 
With equal division of proteins at cell division, the PDF becomes more uniform (red curve).
Filled bars: $\mathrm{Prob} (t \leqs 0.1)$, empty bars: $\mathrm{Prob} (t>0.1)$.
{\bfseries (D)} 
Delay in protein production increases the probability that a pulse occurs just after cell division.
Filled bars: $\mathrm{Prob} (t \leqs 0.1)$, empty bars: $\mathrm{Prob} (t>0.1)$.
{\bfseries (E)}
Mean time between pulses as a function of delay. 
As delay in protein production increases, pulses become significantly less frequent.
We again multiply the birth propensity functions $f_{1}$ and $f_{2}$ by $e^{\gamma \tau }$ for fair comparison (see~\cite{obrien:2012}). 
Other parameters: $\alpha_1=4.5 $, $\alpha_2=12$, $\beta_1=5400$, $\beta_2=600$, $k_1=240$, $k_2=180$, $n=2$, $p=5$, $\delta=0.04$, $\gamma=\ln(2)$.
{\bfseries (F)}
The effects observed in panels~C and~D are robust with respect to modeling variations.
We compare the black PDF from panels~C and~D to PDFs obtained by either varying the volumetric cell division threshold randomly between division events or resampling system parameters upon division.
(See Section~\ref{alternatives} for modeling alternatives.)}
\label{fig:excitable_system}
\end{figure}

We hypothesized that cell cycle noise and delay similarly impact the dynamics of other systems in which escape from metastable states triggers rare events.
We therefore next consider an excitable system with network topology shown in Fig.~\ref{fig:excitable_system}A~\cite{Turcotte14102008}.
The excitable system consists of two genes that code for an activator protein $A$ and a repressor protein $R$. 
The activator activates its own production and that of the repressor, whereas the repressor only inhibits activator activity (by targeting it for degradation).
Thus $R$ acts as a protease.

The excitable system is described by the hybrid framework
\begin{subequations}
\label{e:excitable_network}
\begin{alignat}{2}
&\emptyset \xdashrightarrow[\tau_{1}]{\psi_{1} (X)} X \xrightarrow{\delta XY} \emptyset & &\qquad \text{(simulate using dSSA)}
\\
&\emptyset \xdashrightarrow[\tau_{2}]{\psi_{2} (X)} Y & &\qquad \text{(simulate using dSSA)}
\\
&\frac{\mrm{d} \Omega }{\mrm{d} t} = \gamma \Omega & &\qquad \text{(deterministic cell growth)}
.
\end{alignat}
\end{subequations}
Here $X$ and $Y$ denote the number of molecules of activator $A$ and repressor $R$, respectively.
The quantities $\tau_{1}$ and $\tau_{2}$ are the delay times associated with production.
For simplicity, we assume that $\tau_{1} = \tau_{2} = \tau$.

As before, dashed arrows represent reactions with delay; solid arrows indicate no delay.
The reaction rate $\psi_{i} (\cdot )$ is given by $\psi_{i} (\cdot ) = \Omega (t) f_{i} (\cdot / \Omega (t))$, where $f_{i}$ is the propensity
\begin{equation*}
f_{i} (\cdot ) = \alpha_i +\frac{\beta_i (\cdot )^{n}}{k_i^n + (\cdot )^{n}}.
\end{equation*}

In the $\Omega \to \infty $ limit, system~\eqref{e:excitable_network} is described by the delay differential equations
\begin{subequations}
\label{e:excitable_dRRE}
\begin{align}
\frac{\mrm{d} x}{\mrm{d} t} &= \alpha_1 + 
\frac{\beta_1 x(t - \tau )^{n}}{k_1^n + x(t - \tau )^{n}} - \delta xy - \gamma x
\\
\frac{\mrm{d} y}{\mrm{d} t} &= \alpha_2 + 
\frac{\beta_2 x(t - \tau )^{p}}{k_2^p + x(t - \tau )^{p}} - \gamma y,
\end{align}
\end{subequations}
where $x$ and $y$ denote the concentrations of activator $A$ and repressor $R$, respectively. 

We choose parameters for which system~\eqref{e:excitable_dRRE} has one stable stationary point (with low activator and repressor concentrations), one saddle point, and an unstable spiral point (Fig.~\ref{fig:excitable_system}B).  
In the stochastic ($\Omega < \infty $) regime, the stable stationary point becomes metastable and the system is excitable.
Fluctuations can cause a trajectory to exit the basin of attraction of the metastable point, leading to an excursion around the unstable steady states followed by a return to the basin (Fig.~\ref{fig:excitable_system}B). 
These noise-induced pulses result from interactions between $A$ and $R$: If fluctuations cause activator concentration to increase (or repressor concentration to decrease), positive feedback leads to the production of additional activator and repressor.
Eventually, the repressor (protease) degrades most of the activator, returning the system to the metastable state. 

This system also displays the concentration effect. Pulses most likely occur near the beginning of the cell cycle (Fig.~\ref{fig:excitable_system}C, black PDF), and this effect intensifies with delay (Fig.~\ref{fig:excitable_system}D, green PDF).

As before, the effect is robust with respect to protein number.
The effect is also independent of modeling details:
Fig.~\ref{fig:excitable_system}F shows similar behavior when the cell division threshold varies randomly between divisions, and when we model the effects of division on cellular machinery by resampling system parameters upon division.
(See Section~\ref{alternatives} for modeling alternatives).

The excitable system may be viewed as a bistable system with the two metastable states identified.
Consequently, we hypothesized that increasing delay will increase the mean gap between pulses.
Fig.~\ref{fig:excitable_system}E verifies this prediction.

\section{A three-states model for the concentration effect}
\label{s:symbol}

We introduced a $3$-states reduced model in~\cite{Gupta-stabilization-PRL:2013} to explain why transcriptional delay stabilizes bistable genetic networks in the absence of cell cycle noise.
Here, we extend this $3$-states model to explain the concentration effect.
Although we formulate our extension for bistable switches, similar modeling can be done for excitable systems.

We motivate the extension by intuitively explaining the concentration effect.
At zero delay, transitions between metastable states most likely occur just after cell division because stochastic fluctuations are maximal at the beginning of the cell cycle.
As delay increases, transitions between metastable states due entirely to reaction noise become less frequent~\cite{Gupta-stabilization-PRL:2013}.
Since partitioning noise does not depend on delay, it follows that partitioning noise becomes more important for transitions as delay increases and therefore that the concentration effect intensifies.

\begin{figure}[ht]
\begin{center}
\includegraphics[scale=1.1]{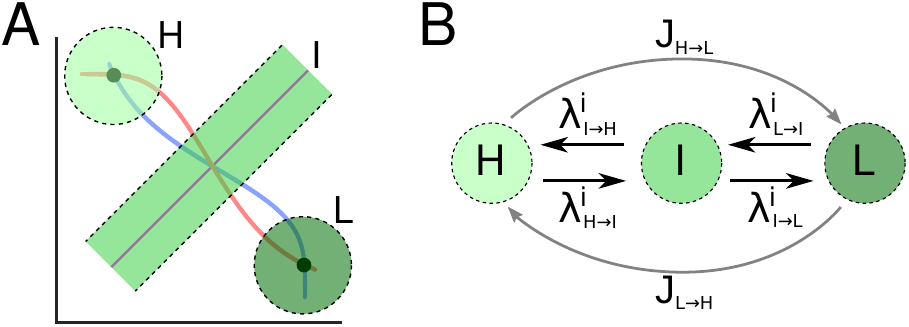}
\end{center}
\caption{$3$-states model for the co-repressive toggle switch.
{\bfseries (A)}
States $H$ and $L$ correspond to disks around the metastable points.
State $I$ corresponds to a tube around the unstable manifold of the saddle point.
{\bfseries (B)}
Continuous-time transition rates model transitions caused by chemical reaction noise (black arrows).
Jumps at cell division times model transitions due to partitioning noise (gray arrows).
}
\label{fig:3states-model}
\end{figure}

Fig.~\ref{fig:3states-model}A shows schematically the three states for the co-repressive toggle switch.
States $H$ and $L$ correspond to neighborhoods of the two metastable states. 
State $I$, an intermediate state, corresponds to a neighborhood of the separatrix between the basins of attraction of the two metastable states.

To capture transitions caused by chemical reaction noise, we introduce continuous-time transition rates (see Fig.~\ref{fig:3states-model}B).
For each pair $j$ and $k$ of adjacent states, let $\lrate{i}{j}{k}$ denote the transition rate from state $j$ to state $k$, given that $\tau $ units of time in the past, the system was in state $i$.
As before, $\tau \geqs 0$ represents transcriptional delay.
The probability of moving from $H$ to $I$ in a time interval $\Delta t$, for example, is given by $\lrate{H}{H}{I} \Delta t$ assuming that $\tau $ units of time in the past the system was in state $H$.
These transition rates are decreasing functions of cell cycle phase.
Since they depend only weakly on the phase, we assume they are constant for simplicity.

To capture transitions caused by partitioning noise, we allow discrete-time jumps at cell division times.
At each such time, a trajectory in state $H$ jumps to $L$ with probability $J_{H \to L}$, while a trajectory in state $L$ jumps to state $H$ with probability $J_{L \to H}$.
Discrete-time jumps into and out of the intermediate state could be added as well, though these are not needed to explain the concentration effect.
Note that while the continuous-time transition rates depend on the past, the jump probabilities do not.

The following assumptions model the dynamics of the toggle switch and imply the concentration effect.
\begin{enumerate}[leftmargin=*, labelindent=\parindent, itemsep=0.4ex, label=\textbf{(A\arabic*)}, ref=A\arabic*]
\item
(Stability)
Each transition rate out of state $I$ ($\lrate{i}{I}{k}$) is at least an order of magnitude larger than all transition rates into $I$ (rates of the form $\lrate{i}{H}{I}$ and $\lrate{i}{L}{I}$).
This assumption forces $H$ and $L$ to function as metastable states.
\item
(Renewal)
The $3$-states model is meant to capture the behavior of the co-repressive toggle when the delay $\tau $ is significantly smaller than mean residence times in the metastable states.
We assume that when the $3$-states system returns to $H$ (or $L$), the system remains in $H$ (or $L$) for at least time $\tau $, so that memory of the history of the trajectory is lost.
\item
\label{i:stickiness}
(Stickiness)
For $i \in \set{ L, I, H }$, define conditional probabilities 
\begin{equation*}
p^{i}_{I \to H} = \frac{\lrate{i}{I}{H}}{\lrate{i}{I}{L} + \lrate{i}{I}{H}}, \qquad p^{i}_{I \to L} = \frac{\lrate{i}{I}{L}}{\lrate{i}{I}{L} + \lrate{i}{I}{H}}.
\end{equation*}
The value $p^{i}_{I \to H}$, for example, is the probability that the system will transition to $H$ rather than $L$, assuming the system is in state $I$ and retains memory of state $i$.
We assume that $p^{H}_{I \to H} > p^{I}_{I \to H}$ and $p^{L}_{I \to L} > p^{I}_{I \to L}$.
This assumption reflects the fact that the birth reaction propensities for the co-repressive toggle depend on the past, not the present.
Consequently, once a trajectory has exited the basin of attraction of a given metastable state, this state will continue to exert a strong pull on the trajectory while the trajectory remembers having been near the metastable state in the past.
\end{enumerate}

The $3$-states model captures the concentration effect.
At zero delay, transitions from $H$ to $L$ concentrate at zero cell cycle phase due to the jump probability $J_{H \to L}$.
As delay increases, transitions from $H$ to $L$ due entirely to reaction noise become less frequent.
Let $\mrm{CMFPT}_{H \to L}$ denote the mean first passage time from $H$ to $L$, conditioned on a discrete-time direct jump from $H$ to $L$ never occurring.
We show in Section~\ref{s:3states-computations} that this conditional mean first passage time has the form
\begin{equation*}
\begin{split}
\mrm{CMFPT}_{H \to L} &\approx (\text{expected number of failed transitions}) \times (\text{mean time of failed transition}) 
\\
&\qquad {}+ (\text{mean time of successful transition}),
\end{split}
\end{equation*}
where a failed transition occurs when the system moves from $H$ to $I$ and then back to $H$ while a successful transition occurs when the system completes the $H \to I \to L$ path.
Further, we show that $\mrm{CMFPT}_{H \to L}$ increases rapidly as a function of $\tau $ because the expected number of failed transitions before a successful transition increases rapidly as a function of $\tau $.
Since the jump probability $J_{H \to L}$ does not depend on $\tau $, it follows that as $\tau $ increases, the fraction of $H$ to $L$ transitions due to direct $H$ to $L$ discrete-time jumps increases.
The concentration effect therefore intensifies.

\section{Cell cycle noise shapes temporal correlations}
\label{s:correlations}

We have shown previously that cell cycle noise shapes the temporal correlations of protein expression in genetic regulatory networks with nontrivial dynamics~\cite{veliz_noise:2015}. 
Fluctuations in parameters or in upstream variables can induce correlations of the dynamics of sister cells after cell division. 
Here we explore this effect further.

\subsection{Constitutive protein production}

As before, we first study constitutive protein production. 
We consider the simplest case in which protein production involves upstream fluctuations: production of a protein $Q$ that depends on a constitutively produced upstream protein $P$. 
The reaction network is given by
\begin{equation}
\label{e:pfeed_production_gfp}
\emptyset \xrightarrow{\psi_1 (X)} P, \ \
\emptyset \xrightarrow{\psi_2 (X,Y)} Q,
\end{equation}
where protein $P$ is constitutively produced, $Q$ can be interpreted as a reporter protein, and $X$ and $Y$ denote the number of molecules of $P$ and $Q$, respectively.
Reaction rates are given by $\psi_1(X)=\Omega f_1(X/\Omega)$ with $f_1(x)=\alpha_1$, and $\psi_2(X,Y)=\Omega f_2(X/\Omega,Y/\Omega)$ with $f_2(x,y)=\alpha_2 x$. 

\begin{figure}[ht]
\begin{center}
\includegraphics[scale=0.4]{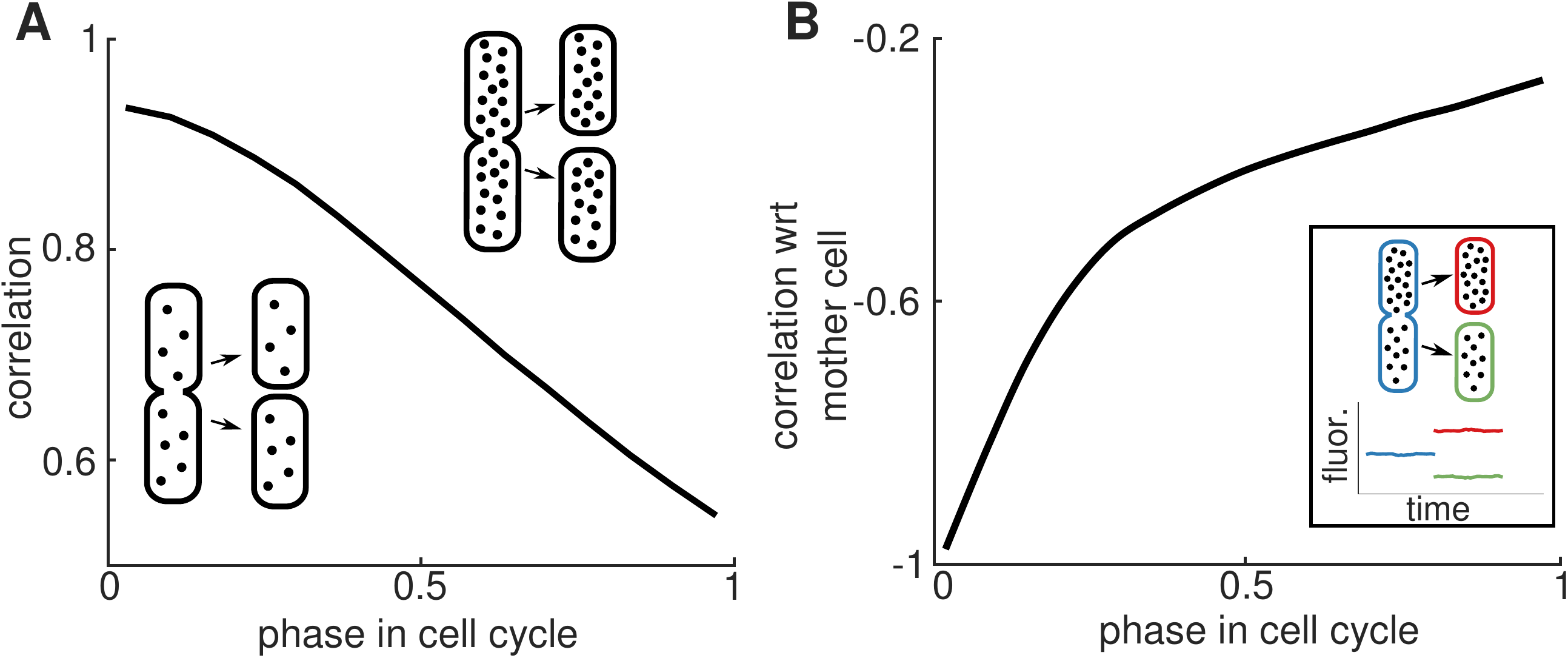}
\end{center}
\caption{Effects of cell cycle noise on correlations for constitutive protein production.
{\bfseries (A)} 
Protein concentrations in daughter cells are highly correlated just after cell division, and decorrelate as the daughter cells grow.
{\bfseries (B)}
Concentrations in daughter cells are anti-correlated with respect to the mother cell.
Due to random partitioning of proteins at cell division, initial protein concentration in one daughter cell will be higher than that of the mother cell (see inset).
Initial concentration in the other daughter cell will be lower than that of the mother cell. 
Parameter values are $\gamma=\ln(2)$, $\alpha_1=10 \gamma$ and $\alpha_2=50 \gamma$, so that mean protein concentrations are $\langle x\rangle =10$ and $\langle y \rangle=500$.
}
\label{fig:stationary}
\end{figure}

Fig.~\ref{fig:stationary}A shows the correlation function  
\begin{equation*}
\rho (y_{1} (t), y_{2} (t)) = \frac{\Cov (y_{1} (t), y_{2} (t))}{\sigma_{y_{1} (t)} \sigma_{y_{2} (t)}},
\end{equation*}
where $y_1$ and $y_2$ are the concentrations of protein $Q$ in the two daughter cells at cell cycle phase $t$.
Daughter cells are highly correlated immediately after cell division.
As the daughter cells grow, their expression levels decorrelate.
We define correlation with respect to the mother cell by
\begin{equation*}
\rho (y_{1} (t) - y_{0}, y_{2} (t) - y_{0}),
\end{equation*}
where $y_{0}$ denotes the concentration of protein Q in the mother cell at the time of cell division.
Fig.~\ref{fig:stationary}B shows that daughter cells are anti-correlated with respect to the mother cell.

\subsection{A synthetic genetic oscillator}

\begin{figure}[ht]
\begin{center}
\includegraphics[scale=0.85]{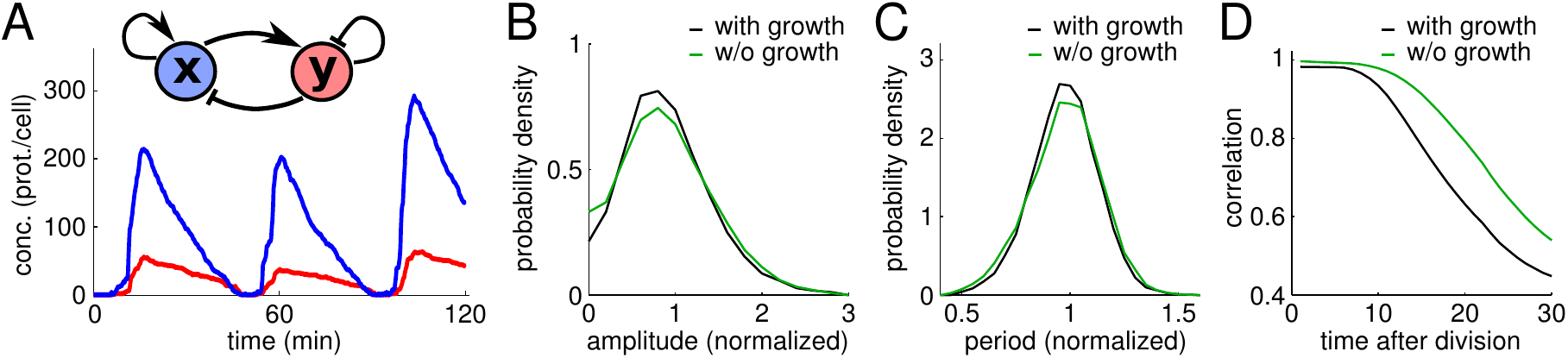}
\end{center}
\caption{Effects of cell growth and division on a synthetic genetic oscillator.
{\bfseries (A)}
Network diagram and traces. 
The stochastic system oscillates with varying period and amplitude.
{\bfseries (BC)}
Probability density functions for amplitude and period. 
Comparing with dilution modeling, explicit cell growth and division modeling reduces the variability in amplitude and period from 0.55 and 0.17 (CV) to 0.52 and 0.15, respectively. Values are normalized so that the mean is 1. 
{\bfseries (D)}
Correlation between sister cells after cell division. 
Comparing with dilution modeling, explicit cell growth and division modeling significantly reduces the correlation between sister cells after division.
Parameters: $\alpha = 0.06$, $k_{1} = 0.08$, $k_{2} = 0.34$, $\beta_{1} = 30$, $\beta_{2} = 6.25$, $\beta_{3} = 60$, $\delta_{1} = 2.4$, $\delta_{2} = 0.8$, $\delta_{3} = 4.8$, $R_{0} = 0.0046$, $\gamma = 0.03 \: \mrm{min}^{-1}$.
Delay values (minutes): $\tau_{1} = 5.5$, $\tau_{2} = 6.0$, $\tau_{3} = 5.0$.
}
\label{fig:dfbo}
\end{figure}

We show that cell cycle noise impacts temporal correlations for a synthetic genetic oscillator~\cite{Stricker2008}. 
The circuit consists of a repressor, an activator, and a reporter protein. 
The activator activates itself and the repressor, while the repressor represses itself and the activator (Fig.~\ref{fig:dfbo}A), thereby forming linked postive and negative feedback loops. 
This system exhibits robust oscillations;
delay plays a key role in the presence and robustness of these oscillations~\cite{Stricker2008, Mather2009}. 
Fig.~\ref{fig:cell_div_modeling}AB shows experimental data, while 
Fig.~\ref{fig:dfbo}A shows simulations of a model of the oscillator.

The oscillator is described by the biochemical reaction network
\begin{subequations}
\label{e:DFBO_network}
\begin{align}
&\emptyset \xdashrightarrow[\tau_{1}]{\psi_{1} (X,Y)} X \xrightarrow{\varphi_1(X,Y,Z)} \emptyset
\\
&\emptyset \xdashrightarrow[\tau_{2}]{\psi_{2} (X,Y)} Y \xrightarrow{\varphi_2(X,Y,Z)} \emptyset 
\\
&\emptyset \xdashrightarrow[\tau_{3}]{\psi_{3} (X,Y)} Z \xrightarrow{\varphi_3(X,Y,Z)} \emptyset
,
\end{align}
\end{subequations}
where $X$, $Y$, and $Z$ denote the number of molecules of activator, repressor, and reporter protein, respectively.
As before, dashed arrows indicate delayed reactions (with delay $\tau_{i}$) and solid arrows indicate reactions without delay.
The production rate $\psi_{i} (X,Y)$ is given by $\psi_{i} (X,Y) = \Omega f_{i} (X/ \Omega,Y/ \Omega )$, where $f_{i}$ is the propensity
\begin{equation*}
f_{i} (x,y) = \beta_i \left(\frac{\alpha + x/k_1}{1 + x/k_1}\right)\frac{1}{(1+y/k_2)^2}.
\end{equation*}
The enzymatic degradation rate $\varphi_i(X,Y,Z)$ is given by $\varphi_i(X,Y,Z)=\Omega g_i(X/ \Omega,Y/ \Omega , Z / \Omega )$, where the propensities $g_{i}$ are given by
\begin{equation*}
g_{1} (x,y,z) =  \frac{\delta_{1} x}{R_0+x+y+z}, \qquad g_{2} (x,y,z) =  \frac{\delta_{2} y}{R_0+x+y+z}, \qquad g_{3} (x,y,z) =  \frac{\delta_{3} z}{R_0+x+y+z}.
\end{equation*}
In the $\Omega \to \infty $ limit, the oscillator is described by the delay reaction rate equations
\begin{subequations}
\label{e:DFBO_dRRE}
\begin{align}
\frac{\mrm{d} x}{\mrm{d} t} &= 
\beta_1 \left(\frac{\alpha + x(t - \tau_{1})/k_1}{1 + x(t - \tau_{1})/k_1}\right)\frac{1}{(1+y(t - \tau_{1})/k_2)^2} - \frac{\delta_1 x}{R_0+x+y+z} - \gamma x
\\
\frac{\mrm{d} y}{\mrm{d} t} &= 
\beta_2 \left(\frac{\alpha + x(t - \tau_{2})/k_1}{1 + x(t - \tau_{2})/k_1}\right)\frac{1}{(1+y(t - \tau_{2})/k_2)^2} - \frac{\delta_2 y}{R_0+x+y+z} - \gamma y
\\
\frac{\mrm{d} z}{\mrm{d} t} &= 
\beta_3 \left(\frac{\alpha + x(t - \tau_{3})/k_1}{1 + x(t - \tau_{3})/k_1}\right)\frac{1}{(1+y(t - \tau_{3})/k_2)^2} - \frac{\delta_3 z}{R_0+x+y+z} - \gamma z.
\end{align}
\end{subequations}

We work with parameters for which system~\eqref{e:DFBO_dRRE} exhibits degrade and fire oscillations~\cite{Mather2009}. 
In the stochastic ($\Omega < \infty $) regime, the system still oscillates (Fig.~\ref{fig:dfbo}A).
We compare statistical properties of the oscillations obtained using explicit cell division modeling with the hybrid algorithm to those obtained using the dilution modeling framework.
(See Section~\ref{s:modeling} for a review of these frameworks.)

Variability in the amplitude and period of the oscillations seems to be insensitive to modeling framework.
We find that the hybrid framework produces a small CV decrease in amplitude and period relative to the dilution framework (Fig.~\ref{fig:dfbo}BC). 

By contrast, the dynamics of sister cells decorrelate significantly faster when cell growth and division are modeled explicitly (Fig.~\ref{fig:dfbo}D).
(Since the dilution framework looks at dynamics within one representative compartment of a single cell that grows indefinitely, we artificially introduce cell division by creating two identical copies of the system state at division times given by integer multiples of $\ln (2) / \gamma $.)

\section{Discussion}

Our results suggest that cell cycle noise shapes the dynamics of systems with metastable states.
Such states play a variety of important functional roles.
Bistable and excitable circuit architectures allow cellular populations to probabilistically change states in response to environmental or internal pressures~\cite{kepler:2001, Eldar-Elowitz-Functional-2010}.
Bistability is essential for the determination of cell fate in multicellular organisms~\cite{hong:2012}, the regulation of cell cycle oscillations during mitosis~\cite{he:2011}, and the maintenance of epigenetic traits in microbes~\cite{ozbudak:2004}.
Excitable architectures enable transient cellular differentiation~\cite{Meeks_2002,Dupin29042003,Suel_2006}.
Transient differentiation into a genetically competent state in \textit{Bacillus subtilis}, for example, is thought to result from excitable dynamics~\cite{Suel_2006, Maamar-2005, Maamar-2007, Suel-2007, Cagatay-2009}.
Cell cycle noise could be important to the dynamics of all of these systems.

Accurately capturing temporal correlations can be important when identifying and quantifying noise sources in a genetic network.
If one overestimates temporal correlations when attempting to match simulation and experimental data by neglecting cell cycle noise, for example, one then risks overestimating the impact of other noise sources.
Recent work with synthetic oscillators demonstrates the value of explicit cell cycle modeling when matching theory and experiments~\cite{veliz_noise:2015}.
Here, the accuracy of estimates of intrinsic and extrinsic noise derived from experimental data depend crucially on explicit modeling of the cell cycle.

Transcriptional delay is central to the production of robust, tunable oscillations in synthetic genetic circuits containing linked positive and negative feedback loops~\cite{Stricker2008, Tigges-2009-Tunable}.
For bistable genetic circuits, delay can induce an analog of stochastic resonance~\cite{Fischer-Imkeller-2005, Fischer-Imkeller-2006}.
Distributed delay (variability in the delay time) can accelerate signaling in transcriptional signaling cascades~\cite{Josic2011}.
The concentration effect may work harmoniously with such delay-induced effects to confer an evolutionary advantage. 
For example, it was shown in~\cite{Gupta-stabilization-PRL:2013} that transcriptional delay stabilizes bistable circuits.
Mean first passage times between bistable states increase dramatically as delay increases.
Consequently, rare events for these systems simultaneously become rarer and increasingly concentrate near the beginning of the cell cycle as transcriptional delay increases.
Delay tuning could therefore function as an evolutionary design principle. 

While transition rates in the $3$-states model can be fit to experimental data, it would be valuable to determine how to compute these rates directly from full models of the dynamics.
This is a challenging large deviations problem because reactions with delay produce non-Markovian dynamics.

While it is but one of many sources of noise in genetic circuits, experimental data suggest that fluctuations induced by cell division significantly impact gene network dynamics~\cite{veliz_noise:2015,Pone_Endres:2015}. 
Such divisions are rarely included explicitly in models, as the effects 
of growth are typically described by simple dilution terms. 
Our results therefore suggest that cell growth and division should be modeled explicitly in order to accurately capture the dynamics of gene circuits.


\section{Supplement: Computations for the three-states model}
\label{s:3states-computations}

We compute the mean first passage time from state $H$ to state $L$, conditioned on a discrete-time direct jump from $H$ to $L$ never occurring.
Let $\mrm{CMFPT}_{H \to L}$ denote this conditional mean first passage time.

Let $f_{H}$ denote the conditional probability that $H \to I \to H$ occurs (the transition attempt fails) given that $H \to I$ has occurred and that the system had been in state $H$ for at least $\tau $ units of time when $H \to I$ occurred.
Let $F_{H}$ denote the corresponding random time needed to complete the $H \to I \to H$ loop given that $H \to I$ has occurred.
Let $P_{F_{H}}$ denote the probability density function for $F_{H}$.
We have
\begin{equation*}
P_{F_{H}} (t) = \frac{1}{f_{H}}
\left\{
\begin{alignedat}{2}
&\lrate{H}{I}{H} \exp \big( - (\lrate{H}{I}{H} + \lrate{H}{I}{L}) t \big) & &(0 < t \leqs \tau )
\\
&\lrate{I}{I}{H} \exp \big( - (\lrate{H}{I}{H} + \lrate{H}{I}{L}) \tau - (\lrate{I}{I}{H} + \lrate{I}{I}{L}) (t - \tau ) \big) \qquad & &(t > \tau )
\end{alignedat}
\right.
\end{equation*}
Integrating over $[0, \infty )$ gives
\begin{equation*}
f_{H} = [1 - Z_{H} (\tau )] p^{H}_{I \to H} + Z_{H} (\tau ) p^{I}_{I \to H},
\end{equation*}
where
\begin{equation*}
Z_{H} (\tau ) = \exp \big( - (\lrate{H}{I}{H} + \lrate{H}{I}{L}) \tau \big).
\end{equation*}
Having computed $f_{H}$, we are in position to estimate the conditional mean first passage time $\mrm{CMFPT}_{H \to L}$. 
Let $S_{H}$ denote the random time needed to complete the $H \to I \to L$ pathway (a successful transition) given that $H \to I$ has occurred.
In terms of $f_{H}$, $F_{H}$, and $S_{H}$,
\begin{equation}
\label{e:mfpt_3states}
\mrm{CMFPT}_{H \to L} \approx \frac{f_{H}}{1 - f_{H}} \left( \mbb{E} [F_{H}] + \frac{1}{\lrate{H}{H}{I}} \right) + \mbb{E} [S_{H}] + \frac{1}{\lrate{H}{H}{I}}.
\end{equation}
The crucial term in~\eqref{e:mfpt_3states} is the factor $f_{H} (1 - f_{H})^{-1}$.
Since $f_{H}$ is a $\tau $-dependent convex combination of $p^{H}_{I \to H}$ and $p^{I}_{I \to H}$ that rapidly transitions from $p^{I}_{I \to H}$ to $p^{H}_{I \to H}$ as $\tau $ increases away from zero, assumption~\pref{i:stickiness} implies that $f_{H} (1 - f_{H})^{-1}$ increases as $\tau $ increases away from zero.
This causes $\mrm{CMFPT}_{H \to L}$ to increase rapidly as $\tau $ increases away from zero.

\section{Supplement: Modeling genetic regulatory networks}
\label{s:modeling}

The hybrid framework we use in this work to model genetic regulatory networks explicitly includes cell growth and division.
This expository section details our hybrid framework and compares it to a well-known modeling hierarchy that uses dilution as a proxy for cell growth and division.

Consider a genetic regulatory system consisting of a single protein $P$ that drives its own production.
The creation of $P$ may be described by the reaction
\begin{equation}
\label{e:pfeed_production}
\emptyset \xdashrightarrow[\mu]{\psi (X)} P,
\end{equation}
where $X$ is the number of molecules of protein $P$, $\psi (X)$ is the reaction rate, and $\mu $ is a probability measure supported on some finite interval, $[0, \tau_{0}],$ that models  the random time from the initiation of transcription to the completion of functional protein (transcriptional delay).
The reaction rate is given by $\psi (X) = \Omega f (X / \Omega )$, where $\Omega $ denotes system volume and $f$ is the propensity function for the reaction.
For example, $f$ may be given by
\begin{equation*}
f(x) = \alpha + \frac{\beta x^{b}}{c^{b} + x^{b}},
\end{equation*}
written in terms of the concentration $x \defas X / \Omega $.

When modeling systems such as~\eqref{e:pfeed_production}, one must decide how to account for cell growth and division. 
One can model cell growth and division explicitly (Fig.~\ref{fig:cell_div_modeling}C,D) or use dilution as a proxy (Fig.~\ref{fig:cell_div_modeling}E,F).
In the second case, one effectively treats the system as a single cell that grows indefinitely and looks at the dynamics within a representative compartment of this single cell (equivalently, within a representative subvolume of the system).
We now describe these two frameworks in detail.
We focus on the reaction described by~\eqref{e:pfeed_production} for the sake of clarity, but our description applies to biochemical reaction networks of any size.

\subsection{Dilution modeling framework}

Genetic regulatory networks (GRNs) can be simulated using an exact delay stochastic simulation algorithm (dSSA) to account for transcriptional delay~\cite{Barrio2006, Bratsun2005, Josic2011, Schlicht_Winkler_2008}.
Since the dSSA models the genetic regulatory network as a stochastic birth-death process and tracks molecule numbers instead of concentrations, dilution is treated by augmenting the GRN with artificial dilution `reactions'.
System~\eqref{e:pfeed_production} in particular takes the augmented form
\begin{equation}
\label{e:pfeed_dilution_fwork}
\emptyset \xdashrightarrow[\mu]{\psi (X)} P \xrightarrow{\gamma X} \emptyset ,
\end{equation}
where $\gamma X$ denotes the rate of the dilution `reaction' and the solid arrow indicates that the dilution `reaction' has no delay associated with it.

The dSSA is implemented as follows for~\eqref{e:pfeed_dilution_fwork}:
Suppose the number of molecules of $P$ and the state of the queue is known at time $t_{0}$.
(The queue accounts for the lag between the initiation of transcription and the production of mature product by storing reactions that have started but are not yet complete.)

\begin{itemize}
\item
Sample a waiting time $t_{w}$ from an exponential distribution with parameter $\psi (X) + \gamma X$.

\item
If there is a reaction in the queue that is scheduled to exit at time $t_{q} < t_{0} + t_{w}$, then advance to time $t_{q}$ and perform the updates $t_{0} \mapsto t_{q}$ and $X \mapsto X+1$.
Finish by sampling a new waiting time for the next reaction.

\item
If no reaction exits the queue before time $t_{0} + t_{w}$, then randomly choose the birth reaction or dilution `reaction' with probabilities proportional to $\psi (X)$ and $\gamma X$, respectively.
If the birth reaction is selected, put this reaction into the queue along with an exit time $t_{q}$, where the reaction completion time $t_{q} - t_{0}$ is sampled from the probability measure $\mu $.
If the dilution `reaction' is selected, perform the update $X \mapsto X-1$.
\end{itemize}

Crucially, the system volume $\Omega $ is treated as a parameter (not as a dynamic variable) in the dilution modeling framework.
It is as if the GRN operates within a single unitary cell that grows forever and the dSSA focuses on a subdomain of volume $\Omega $ (see Fig.~\ref{fig:cell_div_modeling}(EF)).

The dSSA is especially useful when system size is small and stochastic effects are important.
For moderate system sizes, however, the delay chemical Langevin equation (dCLE) offers several advantages.
The dCLE is a stochastic differential equation that models the evolution of the {\itshape concentrations} of the species in the GRN. The dCLE is more computationally efficient than the dSSA at moderate to large numbers of gene transcripts, and can be easier to study analytically.
For~\eqref{e:pfeed_dilution_fwork}, the dCLE is given by
\begin{equation*}
\mrm{d} x_{t} = \left( \int_{0}^{\tau_{0}} f(x_{t-s}) \, \mrm{d} \mu (s) - \gamma x_{t} \right) \mrm{d} t + \frac{1}{\sqrt{\Omega }} \left( \int_{0}^{\tau_{0}} f(x_{t-s}) \, \mrm{d} \mu (s) + \gamma x_{t} \right)^{\frac{1}{2}} \mrm{d} W_{t},
\end{equation*}
where $W_{t}$ denotes one-dimensional Brownian motion.
When system size is large and stochastic effects are unimportant, we may model the GRN using the delay reaction rate equation (dRRE) obtained by taking the $\Omega \to \infty $ limit in the dCLE.
For~\eqref{e:pfeed_dilution_fwork}, the delay reaction rate equation has the form
\begin{equation}
\label{e:pfeed-dRRE}
\frac{\mrm{d} x_{t}}{\mrm{d} t} = \int_{0}^{\tau_{0}} f(x_{t-s}) \, \mrm{d} \mu (s) - \gamma x_{t}.
\end{equation}
Notice the familiar dilution term $- \gamma x_{t}$ in~\eqref{e:pfeed-dRRE}.

We have briefly summarized the modeling hierarchy ($\text{dSSA} \to \text{dCLE} \to \text{dRRE}$) that constitutes the dilution modeling framework.
Higham~\cite{Higham:2008:Modeling} surveys this hierarchy in detail for systems without delay.
For systems with delay, see~\cite{Gupta_jcp2014} for a quantitative mathematical analysis of the relationship between delay birth-death processes and their approximating delay chemical Langevin equations.

\subsection{Explicit cell division modeling}

Cell division is a complex, multi-stage process that has been modeled in detail~\cite{chen:2000, Lu_IEEP_2004,Huh2011}.
We are interested in assessing the impact of cell growth and division on dynamics that occur on timescales much longer than that of cell division itself.
Consequently, we assume that cell division occurs instantaneously.

Several questions must be answered when modeling cell growth and division explicitly.
\begin{itemize}

\item
Does one model cell volume growth deterministically or stochastically?

\item
Assuming cell division is triggered when $\Omega $ reaches a threshold, does one choose a fixed threshold or allow it to vary randomly between cell division events?

\item
How does cell division impact molecular species and GRN parameters?
For instance, are proteins partitioned binomially, or does one include clustering effects?
How does one account for the division of cellular machinery?
\end{itemize}

\begin{figure}[ht]
\begin{center}
\includegraphics[scale=0.9]{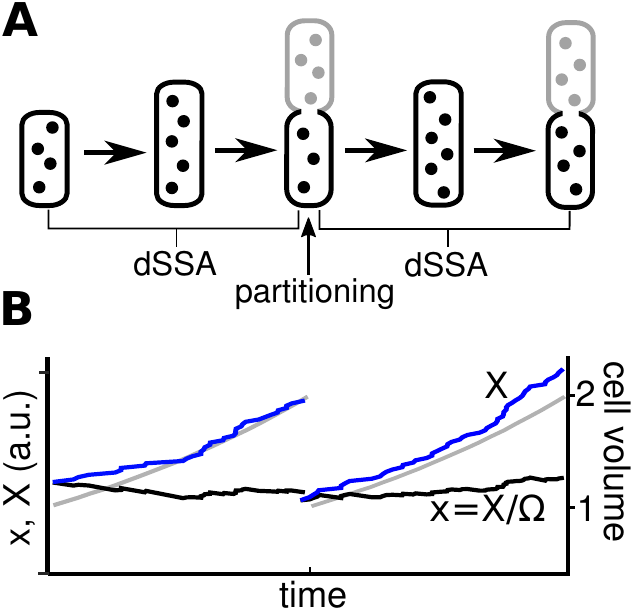}
\end{center}
\caption{Explicit cell growth and division modeling.
{\bfseries (A)} 
Between cell division events, we simulate the reaction network using the dSSA with time-varying volume $\Omega (t)$.
When a cell divides, we partition proteins and the contents of the queues between the two daughter cells. 
We can track the dynamics along one lineage (black cells), or along multiple lineages (black and gray cells).
{\bfseries (B)} 
Sample trajectories for a system in equilibrium.
Cell volume $\Omega (t)$ grows exponentially between cell division events and resets at the moment of division (gray curve).
Protein number $X(t)$ suddenly drops at the moment of cell division due to binomial partitioning (blue curve).
Protein concentration $x(t) = X(t) / \Omega (t)$ is shown in black.
}
\label{fig:dSSA_div}
\end{figure}

In this work, we adopt a hybrid approach that combines the dSSA with deterministic cell volume growth and a fixed volumetric threshold at which cell division is triggered.
For~\eqref{e:pfeed_production} in particular, the hybrid model we use is given by the augmented system
\begin{subequations}
\label{e:pfeed_hybrid}
\begin{alignat}{2}
&\emptyset \xdashrightarrow[\mu]{\psi (X)} P & &\qquad \text{(simulate using dSSA)}
\label{e:pfeed_hybrid-dSSA}
\\
&\frac{\mrm{d} \Omega }{\mrm{d} t} = \gamma \Omega & &\qquad \text{(cell division occurs when $\Omega = \Omega_{\theta }$)}.
\label{e:pfeed_hybrid-volume}
\end{alignat}
\end{subequations}
We simulate hybrid systems such as~\eqref{e:pfeed_hybrid} in the following way:
\begin{itemize}
\item
Simulate the reaction network itself~\eqref{e:pfeed_hybrid-dSSA} using the dSSA.
Notice that the reaction propensities will depend on $\Omega $ in general.
See~\cite{Lu_IEEP_2004} for information about time-dependent dSSAs.
\item
Initialize cell volume $\Omega $ to $1$.
Cell division occurs when $\Omega = \Omega_{\theta }$. 
We set $\Omega_{\theta } = 2$.
\item
At a cell division time, reset $\Omega $ to $1$ and binomially partition both mature protein and the contents of the queues between the two daughter cells. 
\item
Track either a single lineage or multiple lineages (see Fig.~\ref{fig:dSSA_div}A).
\end{itemize}

\subsection{{\bfseries Alternatives.}} \label{alternatives}
There exist many alternatives to the particular hybrid approach that we adopt here.
Within the dSSA-differential equation hybrid framework, one could randomize the time at which cell division occurs by replacing the deterministic volume growth in~\eqref{e:pfeed_hybrid-volume} with a stochastic differential equation, by treating $\Omega_{\theta }$ as a random variable, or by treating cell division as a `reaction' within the dSSA framework.
One can resample system parameters at cell division to model the partitioning of cellular machinery between the daughter cells.

In Fig.~\ref{fig:excitable_system}  we show the results of simulations with a random division threshold and 
parameter resampling at division.  For the random division threshold simulation, we sample the threshold for each division event from the shifted gamma distribution $1.2 + \eta $, where $\eta $ is gamma-distributed with shape $4$ and scale $0.2$, resulting in a mean of $2$ and a CV of $0.2$ for the division threshold.
For the parameter resampling simulation, we resample the production parameters $\alpha_{i}$ and $\beta_{i}$ after each division from normal distributions with coefficient of variation $0.2$ and the same means as in Fig.~\ref{fig:excitable_system}, $\alpha_1=4.5$, $\alpha_2=12$, $\beta_1=5400$, and $\beta_2=600$.
Other parameters are unchanged from Fig.~\ref{fig:excitable_system}.

\begin{acknowledgments}
This work was partially supported by NIH grant 4R01GM104974 (AVC, MRB, KJ, WO), NSF grant DMS 1413437 (CG, WO), and Welch Foundation grant C-1729.
\end{acknowledgments}

\bibliographystyle{siam}
\bibliography{modeling_genetic_networks_lit_v2}

\end{document}